\documentclass[twocolumn, aps, pra, floatfix, floats, reprint, longbibliography]{revtex4-1}
\usepackage{amsmath}
\usepackage{amssymb}
\usepackage{graphicx}
\makeatletter
\providecommand{\tabularnewline}{\\}

\usepackage{tabularx}

\usepackage{multirow}

\makeatother

\begin{document}

\title{Microresonators fabricated from high-kinetic-inductance Aluminum films}

\author{Wenyuan Zhang}
\thanks{These authors contributed equally to this work.} 
\author{K. Kalashnikov}
\thanks{These authors contributed equally to this work.}
\author{Wen-Sen Lu}
\thanks{These authors contributed equally to this work.}
\author{P. Kamenov}
\author{ T. DiNapoli}
\author{M.E. Gershenson}

\affiliation{Department of Physics and Astronomy, Rutgers University, 136
Frelinghuysen Rd., Piscataway, NJ 08854, USA}

\begin{abstract}
We have studied superconducting coplanar-waveguide (CPW) resonators fabricated from disordered (granular) films of Aluminum. Very high kinetic inductance of these films, inherent to disordered materials, allows us to implement ultra-short (200 $\mu$m at a 5GHz resonance frequency) and high-impedance (up to 5 k$\Omega$) half-wavelength resonators. We have shown that the intrinsic losses in these resonators at temperatures $\lesssim 250$ mK are limited by resonator coupling to two-level systems in the environment. The demonstrated internal quality factors are comparable with those for CPW resonators made of conventional superconductors. High kinetic inductance and well-understood losses make these disordered Aluminum resonators promising for a wide range of microwave applications which include kinetic inductance photon detectors and superconducting quantum circuits.
\end{abstract}
\maketitle

\section{Introduction}
The development of novel quantum circuits for information processing requires the implementation of ultra-low-loss microwave resonators with small dimensions \cite{Devoret2013a}. Superconducting resonators have become ubiquitous parts of high-performance superconducting qubits \cite{Paik2011,Barends2013} and kinetic-inductance photon detectors \cite{Zmuidzinas2012}. An important resource for resonator miniaturization is the kinetic inductance of superconductors, $L_K$, which can exceed the magnetic ("geometrical") inductance by orders of magnitude in narrow and thin superconducting films \cite{Tinkham2004}. High kinetic inductance translates into a high impedance $Z$ of the microwave (MW) elements, slow propagation of  electromagnetic waves, and small dimensions of the MW resonators. Ultra-narrow wires and thin films of Nb and NbN \cite{Niepce2018,Zmuidzinas2012}, TiN \cite{Coumou2013}, InOx\cite{Dupre2017,DeGraaf2018}, and granular Al \cite{Rotzinger2017} were studied recently as candidates for high-$L_K$ applications.

Research in high-$L_K$ elements also has an important fundamental aspect. According to the Mattis-Bardeen (MB) theory \cite{Mattis1958}, the kinetic inductance of a thin superconducting film $L_K (T=0)$ is proportional to the resistance of the film in the normal state, $R_N$, and thus increases with disorder. This theory, however, cannot be directly applied to strongly disordered superconductors near the disorder-driven superconductor-to-insulator transition (SIT). Recent theories predict a rapid decrease of the superfluid density near the SIT and the emergence of sub-gap delocalized modes that would result in enhanced dissipation at microwave frequencies \cite{Feigelman2018,Swanson2014}. Thus, the study of the electrodynamics of strongly disordered superconductors may also contribute to a better understanding of the disorder-driven SIT. 

In this Letter, we present a detailed characterization of the half-wavelength microwave resonators fabricated from disordered Aluminum films. Our interest in high-$L_K$ films was stimulated by the possibility of fabrication of superinductors (dissipationless elements with microwave impedance greatly exceeding the resistance quantum $R_Q=h/(2e)^2$ \cite{Annunziata2010,Manucharyan2009,Bell2012}), and the development of superinductor-based protected qubits \cite{Bell2016}. We have fabricated resonators with an impedance $Z$ as high as 5 k$\Omega$, ultra-small dimensions and relatively low losses. The study of the temperature dependences of the resonance frequency $f_r$ and intrinsic quality factor $Q_i$ at different MW excitation levels allowed us to identify resonator coupling to two-level systems in the environment as the primary dissipation mechanism at $T\lesssim 250 $ mK; at higher temperatures the losses can be attributed to thermally excited quasiparticles.

\section{Experimental details}
The standard method for the fabrication of disordered Al  films is the deposition of Al at a reduced oxygen pressure \cite{Deutscher1973,Mui1984}. Such films consist of nanoscale grains ($3-4~$nm in diameter) partially covered by AlO\textsubscript{x}. We have fabricated the films by DC magnetron sputtering of an Al target in the atmosphere of Ar and $\text{O}_2$. Typically, the partial pressures of Ar and $\text{O}_2$ were $5\times 10^{-3}$ mbar and ($3\div7)\times 10^{-5}$ mbar, respectively (the fabrication details are provided in the Supplementary Materials \cite{supplementary}). The films were deposited onto the intrinsic Si substrates at room temperature. By controlling the deposition rate and $\text{O}_2$ pressure, the resistivity of the studied films can be tuned between $10^{-4}~ \Omega \cdot $cm and $10^{-1}~ \Omega \cdot $cm; the parameters of several representative samples are listed in Table~\ref{tab:table_1}. 

The hybrid microcircuits containing the CPW half-wavelength resonators coupled to a CPW transmission line (TL) have been fabricated using e-beam lithography. As the first step, the 50-$\Omega$ TL was fabricated by the e-gun deposition of a 140-nm-thick film of pure Al on a pre-patterned substrate and successive lift-off. The use of pure Al facilitated the impedance matching with the MW set-up and reduced the number of spurious resonances (a large number of these resonances is observed if high-$L_k$ films are used for both the TL and resonator fabrication). After the second e-beam lithography, several half-wavelength disordered Al resonators were fabricated in the openings in the ground plane. Before each metal deposition, reactive ion etching was used to remove the e-beam resist residue from the substrate surface. The width of the central strip of the resonators varied between 0.5 $\mu$m and 10 $\mu$m, and the strip-ground distance  was fixed at $4~\mu$m.

For the resonator characterization at ultra-low temperatures, we used a microwave setup developed for the study of superconducting qubits \cite{supplementary,Bell2012}. The resonators were designed with the resonance frequencies $f_r \approx
 2-4 $~GHz, which allowed us to probe the first three harmonics of the resonators within the setup frequency range $(2\div12)$ GHz. Different resonance frequencies of the resonators enabled multiplexing in the transmission measurements. In order to ensure accurate extraction of the internal quality factor $Q_i$, the resonators were designed with a coupling quality factor $Q_c$ of the same order of magnitude as $Q_i$. 

\section{\label{sec:level1}MICROWAVE CHARACTERIZATION}
The resonators were characterized using a wide range of MW power $P_{MW}$, two-tone (pump-probe) measurements, and time domain measurements. The resonator parameters $f_r$, $Q_i$, and $Q_c$ were found from the simultaneous measurements of the amplitude and the phase of the transmitted signal $S_{21}(f)$ using the procedure described in Refs. \cite{Khalil2012,Probst2015} and Supplementary Materials \cite{supplementary}. The kinetic inductance $L_K$ of the central conductor of the resonators, which exceeded the magnetic inductance by several orders of magnitude, was calculated as $L_K=1/4f_r^2C$ (the capacitance $C$ between the resonator strip and the ground was obtained in the Sonnet simulations). The parameters of several representative resonators are listed in Table \ref{tab:table_1}.

The measured sheet kinetic inductance $L_{K\Box} \approx 2~\text{nH}/\Box$  is similar to that reported for granular Al films in Ref.$~$\cite{Grunhaupt2018} and TiN in Ref. \cite{Peltonen2017}, and exceeds by a factor-of-2 $L_{K\Box}$ realized for ultra-thin disordered films of InOx \cite{Astafiev2012,Dupre2017}. For the disordered Al films with $\rho<10~ m\Omega \cdot $cm, $L_{K\Box}$ is in good agreement with the result of the MB theory \cite{Mattis1958}, $L_{K\Box} (T=0)=\hbar R_\Box/\pi \Delta(0)$, where  $\Delta(0)$ is the BCS energy gap at $T=0~K$. Deviations from this behavior, observed for highly disordered film (e.g., resonator \#1), will be discussed in a seperate paper \cite{unpublished2018}. Very large values of  $L_{K\Box}$ allowed us to realize the characteristic impedance $Z=\sqrt{L_K/C}$ as high as 5 k$\Omega$ for the resonators with narrow ($ w = 0.7~\mu $m) central strips. The speed of propagation of the electromagnetic waves in such resonators does not exceed 1\% of the speed of light in free space; accordingly, their length is two orders of magnitude smaller than that for the conventional CPW resonators with the impedance $Z=50~\Omega$.

To identify the physical mechanisms of losses in the resonators, we measured the dependences of $f_r$ and $Q_i$ on the temperature ($T=25 \div 450$ mK) and the microwave power $P_{MW}$. Below we show that in the case of moderately disordered films (resonators $\#2-4$), both the dissipation and dispersion at $T<0.25~$ K can be attributed to the resonator coupling to the two-level systems (TLS) in the environment, whereas at higher temperatures they are controlled by the $T$ dependence of the complex conductivity of superconductors, $\sigma(T)=\sigma_1 (T)-i\sigma_2 (T)$ \cite{Mattis1958}.

\begin{table}[htp]
\begin{centering}
\caption{Summary of the measured parameters of AlOx resonators}
\label{tab:table_1} 
\par\end{centering}
\newcolumntype{Y}{>{\centering\arraybackslash}X}
\begin{tabularx}{\linewidth}{ YYYYYYYYY  }
        \hline
        \hline
         \multirow{2}{*}{\#}   & $w$,   & $l$,  & $f_r$,  & $\rho$, & $T_c $, & $L_K $, & Z,    \tabularnewline
         
                              \ &  $\mu$m  &  $\mu$m & GHz &  m$\Omega\cdot$cm &  K & $ {\text{nH}}\Big/{\Box}$ & $k\Omega$   \tabularnewline

         \hline
         1& 11.0& 1090 & 2.42 & 19.2 & 1.4 & 2.0 & 0.6  \tabularnewline
         \hline
         2& 7.4& 765 & 4.05& 4.2 &1.7 & 1.2 &  1.1   \tabularnewline
         \hline
         3& 1.4& 445 & 3.69& 4.2 & 1.7&1.2  &2.9  \tabularnewline
         \hline
         4& 0.7& 265 & 3.88& 9.9 & 1.75 & 2.0 & 5.0   \tabularnewline
         \hline
         \hline
    \end{tabularx}
\end{table}

\begin{figure*}[ht]
\centering
\includegraphics[width=\textwidth]{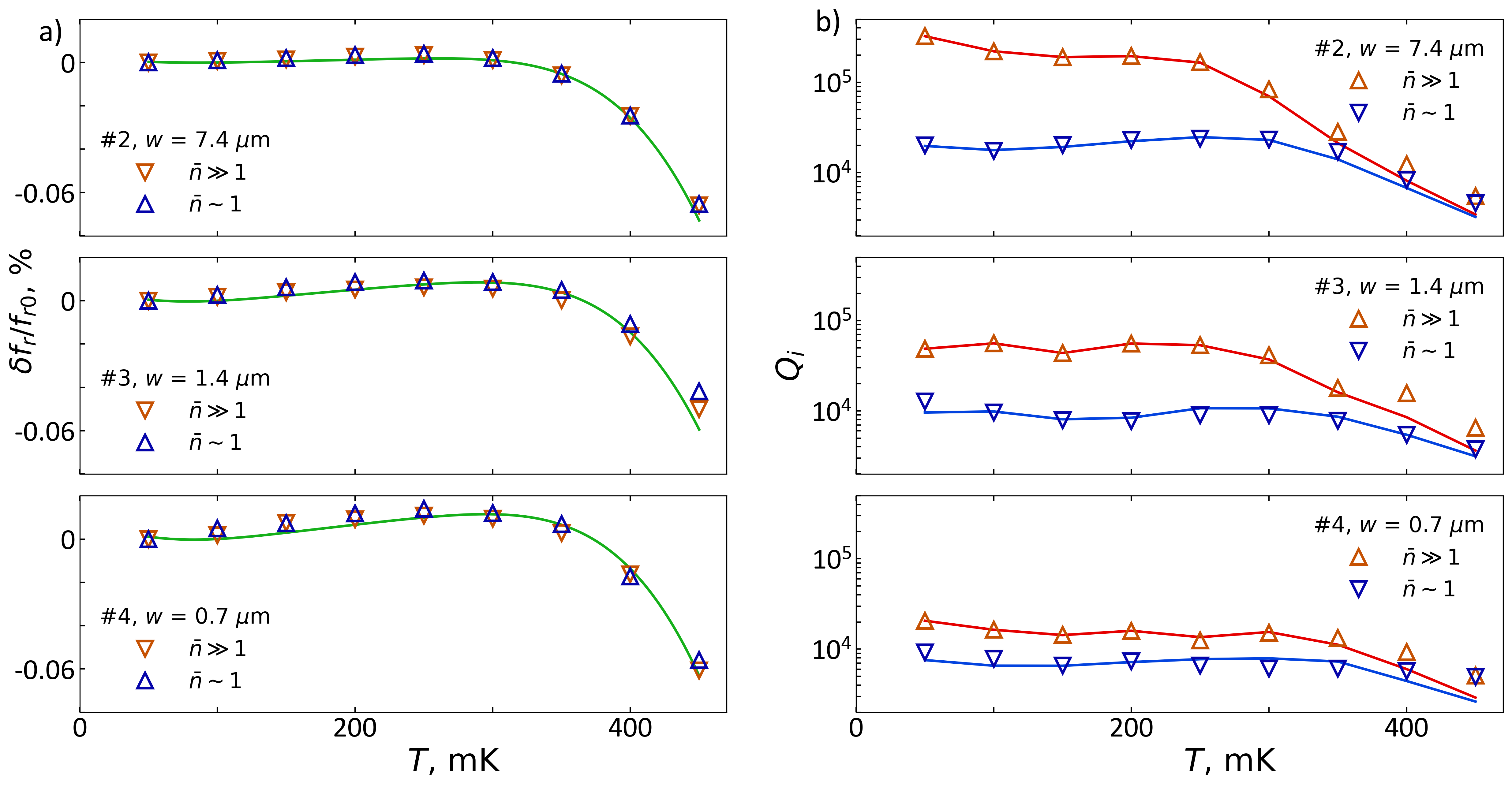}
\caption{ The temperature dependences of resonance frequency shift $\delta f_r^{TLS}(T)/f_{r0} $ (a) and the internal quality factor $Q_i$ (b) for the resonators \#$2-4$ measured at $\bar{n} \approx 1 (\bigtriangledown)$ and $\bar{n} \gg 1 (\bigtriangleup)$. The fitting curves correspond to Eq. (\ref{eq:eq2}) and Eq. (\ref{eq:eq7}), respectively.}
\label{fig:fig1}
\end{figure*}

\subsection{The resonance frequency analysis}

We start the data analysis with the temperature dependence of the relative shift of the resonance frequency $\delta f_r (T)/f_{r0} \equiv [f_r(T)-f_r (25mK)]/f_r (25mK) $. Figure  \ref{fig:fig1}(a) shows the dependences $\delta f_r(T)/f_{r0}$ measured for three resonators (\#$2-4$) with different width $w$. The low-temperature part of $\delta f_r(T)/f_{r0}$ is governed by the $T$-dependent TLS contribution to the imaginary part of the complex dielectric permittivity $\epsilon(T)=\epsilon_1 (T)+i\epsilon_2 (T)$. It should be noted that, in contrast to the TLS-related losses, the frequency shift $\delta f_r^{TLS}$ is expected to be weakly power-dependent \cite{Gao2008}. Indeed, the temperature dependences measured for the different values of $P_{MW}$ almost coincide; this simplifies the analysis and reduces the number of fitting parameters. The low-temperature part of $\delta f_r^{TLS}(T)$ is well described by the following equation \cite{Zmuidzinas2012}:
\begin{align}
    \frac{\delta f_r^{TLS}(T)}{f_{r0}} = \frac{V_f \delta_0}{\pi} \Bigg[  \Psi_{\Re} \bigg(\frac{1}{2}+\frac{1}{2\pi i} \frac{h f_r}{k_{B}T} \bigg) 
    -\ln \bigg(\frac{h f_r}{k_{B} T} \bigg) \Bigg].
    \label{eq:eq1}
\end{align}
Here $\Psi_{\Re}(x)$ is the real part of the complex digamma function, the TLS participation ratio $V_f$ is the energy stored in the TLS-occupied volume normalized by the total energy in the resonator, and the loss tangent $\delta_0$ characterizes the TLS-induced microwave loss in weak electric fields at low temperatures $k_{B} T\ll h f_r$. The product $V_f \delta_0$ is the only fitting parameter, its values are listed in Table \ref{tab:table_2}. The obtained values of $V_f \delta_0$ are close to that found for Al-based \cite{Gao2008} and AlOx-based resonators \cite{Peltonen2017,Wenner2011}. Note that resonator \#4 demonstrates the most pronounced increase of $f_r(T)$ with temperature due to the stronger electric fields and a larger participation ratio characteristic of the high-$Z$ resonators \cite{Sage2011}. 

At $T>0.25$ K, $f_r$ rapidly drops due to the decrease of the superfluid density. The dependences $\delta f_r(T)$ over the whole studied $T$ range can be described as 
\begin{align}
    \delta f_r(T)/f_{r0} = \delta f_r^{TLS}(T)/f_{r0}  + \delta f_r^{MB} (T)/f_{r0}
\label{eq:eq2}
\end{align}
where 
\begin{align}
\frac{\delta f_r^{MB}(T)}{f_{r0}}=\frac{1}{2} \bigg[ \frac{\sigma_2(T)-\sigma_2 (25mK)}{\sigma_2 (25mK)} \bigg]
\label{eq:eq3}
\end{align}
is the resonance shift due to the $T$-induced break of Cooper pairs and subsequent increase of the kinetic inductance, calculated in the thin film limit \cite{Gao2008}. The only free parameter in $\delta f_r^{MB} (T)/f_{r0}$ is the gap energy $\Delta(0)$, which can be found by fitting of the high-$T$ portion of $\delta f_r(T)/f_{r0}$ [Eq. (\ref{eq:eq2})]; the measured ratio $\Delta(0)/T_c$ is about 10\% greater than the BCS value of $1.76k_B$, which is consistent with previously reported data \cite{Pracht2016}. 
\subsection{The quality factor analysis}
\begin{figure}
    \centering
    \includegraphics[width=\linewidth]{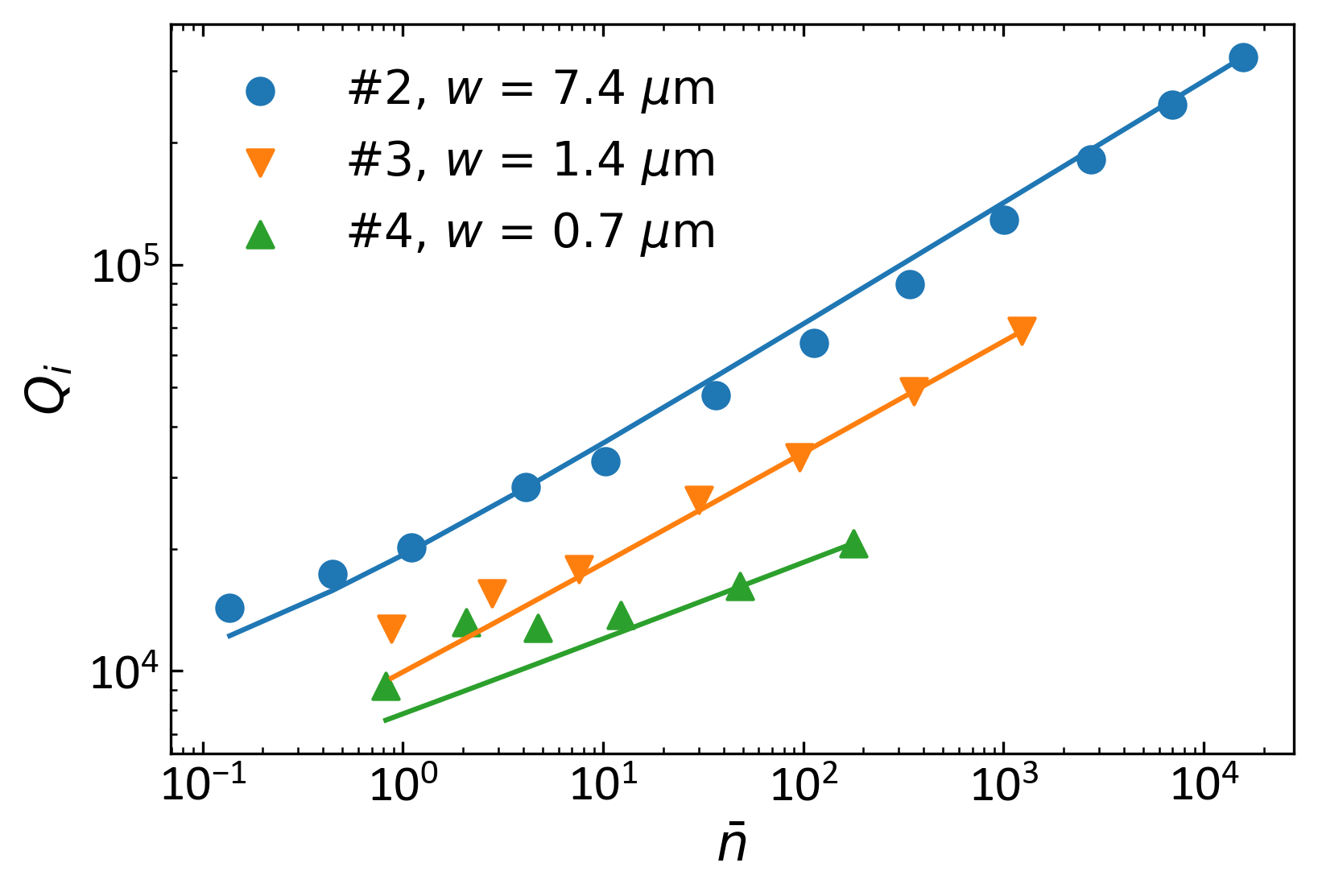}
    \caption{
    The dependences $Q_i(\bar{n})$ at $T\approx 25$ mK for the resonators with different widths. Solid curves represent the theoretical fits of the quality factor governed by TLS losses [Eq. (\ref{eq:eq5}), see the text for details].
    }
    \label{fig:fig2}
\end{figure}

We now proceed with the analyses of losses. We observed the enhancement of the internal quality factor $Q_i$ with increasing the average number of photons in the resonators, $\bar{n} = 2 P_{MW} Q_l^2/(Q_c h f_r ^2 )$ \cite{Bruno2015}, where $Q_l = (1/Q_i +1/Q_C)^{-1}$ is the loaded quality factor. The dependences $Q_i(\bar{n})$ for three resonators with different $w$ measured at the base temperature $\approx 25$ mK are shown in Fig. \ref{fig:fig2}. Similar behavior of $Q_i(\bar{n})$ have been observed for many types of CPW superconducting resonators (see, e.g. \cite{Zmuidzinas2012, Megrant2012} and references therein), including the resonators based on disordered Al films \cite{Rotzinger2017,Grunhaupt2018}. Note that the increase of $Q_i$ with the input MW power $P_{MW}$ is limited by the resonance distortion by bifurcation at $P_{MW}>P_{*}$. For the resonators with $Q_l \gtrsim 10^4$ the onset of bifurcation is observed for the microwave currents $I_*=\sqrt{2P_*/Z}$ which scale approximately as $I_{dp}/\sqrt{Q_l}$ \cite{Swenson2013a}, where $I_{dp}$  is the Ginzburg-Landau depairing current in the central strip \cite{supplementary}.

The power-dependent intrinsic losses can be attributed to the resonator coupling to the TLS with the Lorentzian-shaped distribution
\begin{align}
    g(E_{TLS})\sim \frac{1}{(E_{TLS}-hf_r)^2+(\hbar/\tau_2)^2 },
    \label{eq:eq4}
\end{align}
where $E_{TLS}$ is the energy of TLS and $\tau_2$ is its dephasing time \cite{Pappas2011}. Once the MW power $P_{MW}$ reaches some characteristic level $P_c$ and the Rabi frequency of the driven TLS $\Omega_R \sim \sqrt{P_{MW}}$ exceeds the relaxation rate $1/\sqrt{\tau_1 \tau_2}$, the population of the excited TLS increases, and the amount of energy that the TLS with $f_{TLS} \approx f_r$ can absorb from the resonator decreases. Thus, the high $P_{MW}$ ''burns the hole'' in the density of states (DoS) of dissipative TLS. The width of the ''hole'' is $\kappa/ 2\pi\tau_2$, the power-dependent factor can be written as
\begin{align}
    {\kappa = \sqrt{1+\bigg(\frac{\bar{n}}{n_c}\bigg)^\beta}},
    \label{eq:eq5}
\end{align}
where $\bar{n}$ and ${n_c}$ correspond to $P_{MW}$ and $P_c$, respectively. Note that the exponent $\beta$ is known to be dependent on the electric field distribution in a resonator \cite{Wang2009}, and the characteristic power $n_c$ increases with temperature by orders of magnitude due to a strong $T$-dependence of $\tau_1$ and $\tau_2$ \cite{Goetz2016}. Taking into account the TLS saturation at high temperature, the power dependence of the TLS-related part of the loss tangent can be expressed as follows \cite{Sage2011}:
\begin{align}
    \delta_{TLS}(\bar{n},T)=  \frac{V_f \delta_0}{\kappa} \tanh\bigg(\frac{hf_r}{2k_B T}\bigg). 
    \label{eq:eq6}
\end{align}

 By fitting the experimental data with Eq. (\ref{eq:eq6}) we found $\beta$ and $n_c$, the obtained parameters are listed in Table \ref{tab:table_2}. We found that larger values of $\beta$ correspond to wide strips, and the extracted $n_c$ scales as the square of the electric field on the surface of the resonator. The details of the fitting procedure can be found in Supplementary Materials \cite{supplementary}.

The experimental dependences $Q_i (T)$  measured for resonators $\#2-4$ at $\bar{n} \backsimeq 1$ and $\bar{n} \gg 1$ [Fig. \ref{fig:fig1}(b)] are well described by the sum of the TLS contribution [Eq. (\ref{eq:eq6})] and the MB term $\delta_{MB}=\sigma_1 (T)/\sigma_2(T)$ \cite{Gao2008}: 
\begin{align}
    Q_i (T)=\{\delta_{TLS}(T, \beta, n_c, V_f \delta_0) + \delta_{MB} [T, \Delta(0)] \}^{-1} 	\label{eq:eq7}.
\end{align}
The agreement of measured $Q_i$ with the prediction of Eq. (\ref{eq:eq7}) over the whole measured temperature range proves that the losses in the developed resonators are limited by the sum of TLS and MB terms.

\subsection{The two-tone and time-domain measurements}
\begin{figure*}
    \centering
    \includegraphics[width=\textwidth]{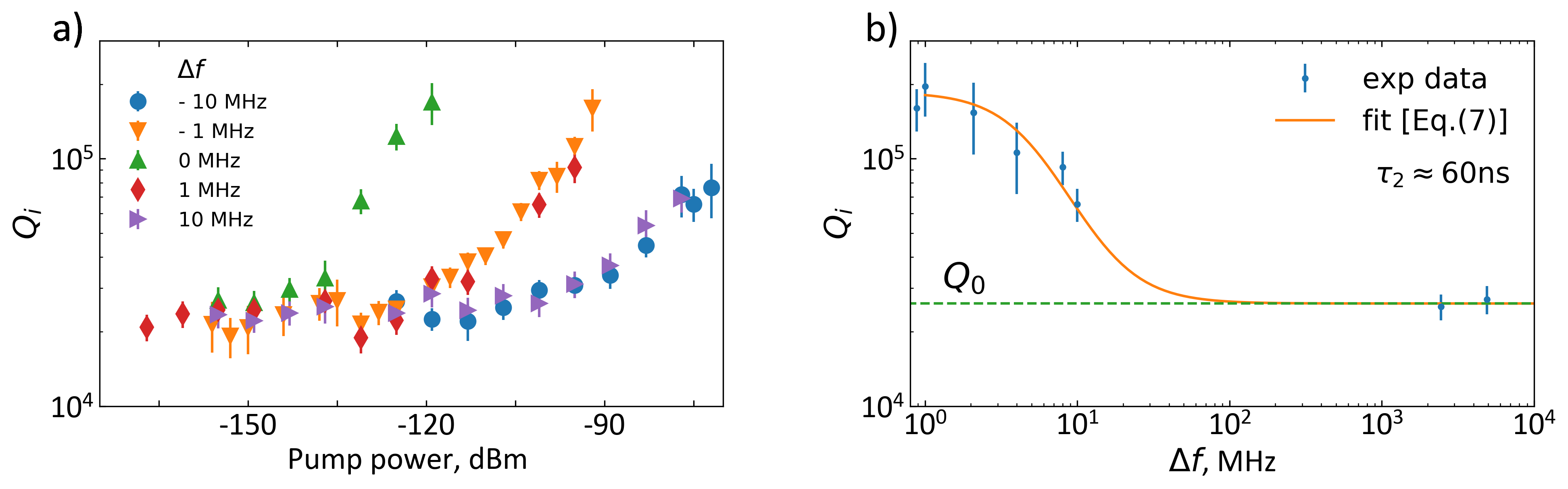}
    \caption{\label{fig:fig3}(a) The dependences of $Q_i$   for resonator \#1 on the pump tone power $P_p$ for several values of detuning $\Delta f$ between resonance and pump frequencies. (b) The values of $Q_i$   measured versus detuning $\Delta f$ at a fixed number of the pump tone photons in the resonator $\bar{n}_p \approx 1000$. }
\end{figure*}

We obtained an additional information on the TLS-related dissipation by performing the pump-probe experiments in which $Q_i$ was measured at a low-power ($n\backsimeq
1$) probe signal while the power $P_{p}$  of the pump signal at the frequency $f_{p}$  was varied over a wide range. Figure \ref{fig:fig3}(a) shows the dependences $Q_i(P_{p} )$ measured at different detuning values $\Delta f = f_{p}-f_r=0$, $\pm1$ MHz, and $\pm10$ MHz. Note that we have not observed any changes in $Q_i$ when the pump signal was applied at the second and third harmonics of the resonator. Also, $Q_i$ was $P_{p}$-independent when we monitored the second harmonic and applied the pump signal at the first harmonic.

Since the resonator coupling to the pump signal varies by several orders of magnitude within the detuning range $0\div10$ MHz, it is more informative to analyze $Q_i$ as a function of the average number of the "pump" photons in the resonator, $\bar{n}_{p}=P_{p}(1-|S_{21}(f_{p} )|^2-|S_{11} (f_{p})|^2)/hf_{p}^2$, where $S_{21}$  and $S_{11}=1-S_{21}$ are the transmission and reflection amplitudes at the pump frequency, respectively. The dependence $Q_i$ on the detuning $\Delta f$ for a fixed $\bar{n}_{p}\approx1000$ is depicted in Fig. \ref{fig:fig3}(b). The resonance behavior of $Q_i (\Delta f)$ is expected since only a narrow TLS band [Eq. (\ref{eq:eq4})] contributes to dissipation: the "hole" extension in the DoS is limited by  $\sim\kappa/\tau_2$ around the pump frequency. Indeed, using the approach developed in \cite{Capelle2018}, one can obtain the following expression:

\begin{table}[t]
\begin{centering}
\caption{Summary of the fitting parameters}
\label{tab:table_2} 
\par\end{centering}
\newcolumntype{Y}{>{\centering\arraybackslash}X}
\begin{tabularx}{\linewidth}{ YYYYY  }
        \hline
        \hline
         \# & ${\Delta(0) }/{ k_B T_c}$  & $\beta$ & $V_f \delta_0 \cdot 10^{-4}$ & $n_c(0)\cdot 10^{-3}$  \tabularnewline
         \hline
          2 & 1.96 & 0.60 & 1.4 & 50 \tabularnewline
         \hline
          3 & 1.98 & 0.55 & 4.8 & 1.6\tabularnewline
         \hline
          4 & 1.88 & 0.38 & 6.7 & 0.23 \tabularnewline
         \hline
         \hline
    \end{tabularx}
\end{table}

\begin{align}
    Q_i (\Delta) = Q_0\bigg[1 + \frac{(\kappa/2\pi\tau_2)^2}{\Delta f^2 + \kappa(1/2\pi\tau_2)^2}\bigg], 	\label{eq:eq8}
\end{align}

where $Q_0$ is the off-resonance quality factor, and introduced by Eq. (\ref{eq:eq5}) factor $\kappa$ might be calculated as $\kappa = Q_{max}/Q_0$. The dephasing time is the only fitting parameter and it is found to be $\tau_2\approx60$ ns. This result agrees with the measurements of the dephasing time for individual TLS in amorphous $Al_2O_3$ tunnel barrier in Josephson junctions \cite{Shalibo2010}.

By application of the MW pulses at the pump frequency, we observed that the characteristic time at which $Q_i$ varies with $P_{p}$ does not exceed 36 ms (see Supplementary Materials \cite{supplementary} for details). For several resonators we have observed the telegraph noise in the resonance frequency on the time scale of $1-10$ s. This noise can be attributed to interactions of the resonators with a small number of strongly coupled TLS.

\section{Summary}
In conclusion, we have fabricated CPW half-wavelength resonators made of strongly disordered Al films. Because of the very high kinetic inductance of these films, we were able to significantly reduce the length of these resonators, down to $\sim 1$\% of that of conventional CPW resonators with a 50 $\Omega$ impedance. Due to ultra-small dimensions and relatively low losses at mK temperatures, these resonators are promising for the use in quantum superconducting circuits operating at ultra-low temperatures, especially for the applications that require numerous resonators, such as multi-pixel MKIDs \cite{Zmuidzinas2012, Swenson2013a}. The high impedance $Z=\sqrt{L_K/C}$ of the narrow resonators can be used for effective coupling of spin qubits \cite{Samkharadze2016, Stockklauser2017}.
The high resonator impedance imposes limitations on the strength of resonator coupling to the transmission line. For the studied CPW resonators with $Z\sim 5~k\Omega$, the strongest realized coupling (when half of the resonator length was used as the element of capacitive coupling to the transmission line) corresponded to $Q_c \sim 10^4$. On the other hand, for many applications, such as large MKID arrays that require a high loaded $Q$ factor, this should not be a limitation.

We have shown that the main source of losses in these resonators at $T\ll T_c$ is the coupling to the resonant two-level systems. A comparison of our results with those of the other groups shows that the obtained $Q_i$ values, increasing from $(1\div2)\times 10^4$  in the single-photon regime to 3$\times10^5$ at high microwave power, are typical for the CPW superconducting resonators with similar TLS participation ratios. This implies that the disorder in Al films does not introduce any additional, anomalous losses. Most likely, the relevant TLS are located near the edges of the central resonator strip either in the native oxide on the Si substrate surface or in the amorphous oxide covering the films. Further increase of $Q_i$   can be achieved by the methods aimed at the reduction of surface participation, such as substrate trenching (see \cite{Calusine2018a} and references within) and increasing the gap between the center conductor and the ground plane \cite{Wang2009}. The evidence for that was provided by the results of Ref. \cite{Grunhaupt2018} obtained for the modified three-dimensional microstrip structures based on disordered Al films. It is also worth mentioning that the losses can be reduced using TLS saturation by the microwave signal outside of the resonator bandwidth but within the TLS spectral diffusion range.
A fundamental issue pertinent to all strongly disordered superconductors is the development of a better understanding of the impedance of superconductors near the disorder-driven SIT. This issue requires further research, and the microwave experiments with the resonators made of disordered Al and other disordered materials demonstrating the SIT may shed light on the nature of this quantum transition.

\begin{acknowledgments}
This work was supported by the NSF award 1708954 and the ARO award W911NF-17-C-0024.
 
\end{acknowledgments}

\section*{Supplementary Materials}

\renewcommand{\theequation}{\arabic{equation}SM}
\renewcommand{\thetable}{S\arabic{table}}
\renewcommand{\thefigure}{S\arabic{figure}}   
\setcounter{figure}{0}
\setcounter{section}{0}
\renewcommand{\bibnumfmt}[1]{[S#1]}
\renewcommand{\citenumfont}[1]{S#1}

\section{Fabrication of microwave resonators}
\begin{figure}[hb]
    \centering
    \includegraphics[width=\linewidth]{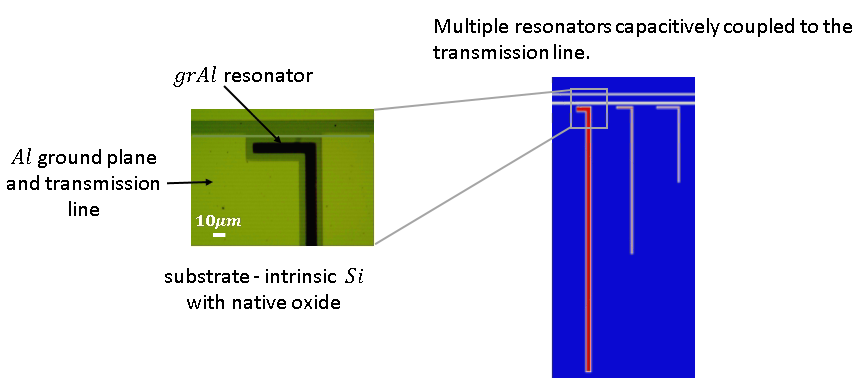}
    \caption{(a) Microphotograph of a portion of the half-wavelength resonator capacitively coupled to the coplanar waveguide transmission line. Light green - Al ground plane and the central conductor of the transmission line, green - silicon substrate, black - the central strip of the resonator made of strongly disordered Al. (b) Several resonators with different resonance frequencies coupled to the transmission line.}
    \label{figS:1}
\end{figure}

All microwave (MW) resonators studied in this work consisted of two parts. First, the 50-Ohm coplanar MW transmission line was formed on an intrinsic Si substrate by electron beam deposition of a 140-nm-thick film of pure Al through a lift-off mask, which comprised of a 300-nm-thick e-beam resist (the top layer) and a 150-nm-thick copolymer (the bottom layer). After the deposition of the bilayer resist and its patterning with e-beam lithography, the sample was placed in a reactive ion etching system and etched with 75 mbar $O_2$ plasma at a power of 30 watts for 30 seconds to remove any resist residue from the substrate surface. The use of this pure Al transmission line facilitated the impedance matching with the MW tract and eliminated spurious resonances. After the second e-beam lithography with alignment precision better than 0.5 $\mu$m, several half-wavelength disordered Al resonators were fabricated on the same substrate by reactive DC magnetron sputtering in a vacuum system with the base pressure of $<1\times10^{-6}$mbar (Fig. \ref{figS:1}). The disordered films were deposited by sputtering of a 6N-purity Al target onto Si substrates held at room temperature. In order to improve reproducibility, prior to the disordered Al deposition the target was pre-cleaned in a pure Ar-plasma by sputtering at a rate of 0.6 nm/s for 5 minutes. The reactive DC sputtering of disordered Al was then initiated by introducing 1 sccm $O_2$ and 115 sccm Ar gas mixture from two independent feedback-controlled mass flow meters (MicroTrak\textsuperscript{TM} and SmartTrak\textsuperscript{TM}). During the sputtering process, typical partial pressures of Ar and $O_2$ were $5\times 10^{-3}$ mbar and ($3\div7)\times 10^{-5}$ mbar, respectively. After the second lift-off, the chip was installed in the sample holder by wire bonding.

\section{Measurement setup}
\subsection{Microwave setup}
\begin{figure}[h]
    \centering
    \includegraphics[width=0.9\linewidth]{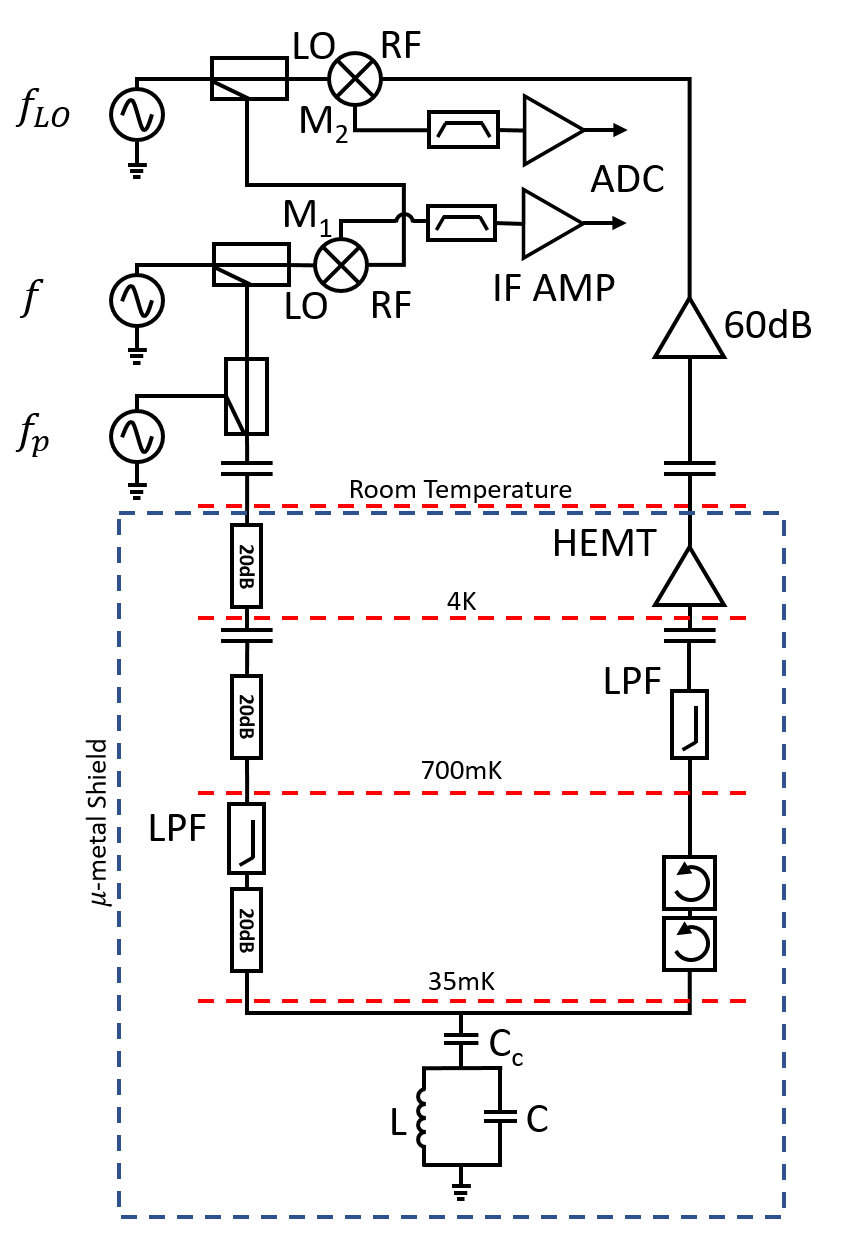}
    \caption{Schematics of the resonator measurement setup}
    \label{figS:figS_uw_setup}
\end{figure}

All measurements were performed in the BlueFors\textsuperscript{TM} BF-SD250 dilution refrigerator with a base temperature of $\sim$25mK. To reduce stray magnetic fields, a $\mu$-metal shield was installed outside of the cryostat.  We used the microwave measurement setup (Fig. \ref{figS:figS_uw_setup}) developed for the research in superconducting qubits; it was described in our previous publication \cite{Bell2012b}. The setup enabled the resonator testing over a wide range of MW power, including the single-photon population regime, the two-tone (pump-probe) and time domain measurements.

The probe signal at $f$ and the pump signal at $f_p$, generated by two microwave synthesizers, were coupled to the input of the cryostat through directional couplers. Depending on the experiment performed, the pump signal could be pulsed using an internal RF switch of the microwave synthesizer. Attenuators and low-pass filters were installed in the microwave input line to prevent leakage of thermal radiation into the resonator.  The signal, after passing the sample, was amplified by a cryogenic high-electron mobility transistor (HEMT amplifier Caltech CITCRYO 1-12, 35 dB gain between 1 and 12 GHz) and two 30dB room-temperature amplifier. Two cryogenic Pamtech isolators (each provides 18dB isolation between 3 and 12 GHz) were anchored at the base temperature to reduce the 5K noise from the HEMT amplifier.  The amplified signal was downconverted to the intermediate frequency (IF) $f_{IF}=|f-f_{LO}|\approx30MHz$ using mixer M1 with the local oscillator signal $f_{LO}$. The IF signal was digitized using the card AlazarTech ATS 9870 at 1GS/s. The magnitude and phase of the signal $S_{21}$ was obtained by digital demodulation as $a=\sqrt{(\langle a^2 (t)  \sin^2(2\pi f t)\rangle + \langle a^2 (t)  \cos^2(2\pi f t)\rangle)}$ and $\phi= \arctan (\langle a^2 (t)  \sin^2(2\pi f t)\rangle/\langle a^2 (t)  \cos^2(2\pi f t)\rangle ) -\phi_0$ (here $\langle ... \rangle$ stands for the time averaging over integer number of periods, typically $10^6$). The reference phase $\phi_0$ was provided by mixer M2. 

\begin{figure}[h]
    \centering
    \includegraphics[width=\linewidth]{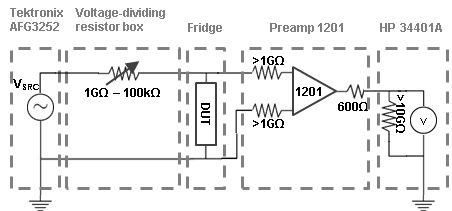}
    \caption{Schematics of the DC setup}
    \label{figS:dc_electronics}
\end{figure}

\subsection{DC setup}
On the same resonator chip, we also patterned Hall bars to measure critical currents for the disordered Aluminum films. The critical currents were measured using an Arbitrary Waveform  Generator (Tektronix AFG3252)  and HP 34401A multimeter (see Fig. \ref{figS:dc_electronics}). 


\section{The procedure of extracting the quality factors}
\begin{figure}[ht]
    \centering
    \includegraphics[width=\linewidth]{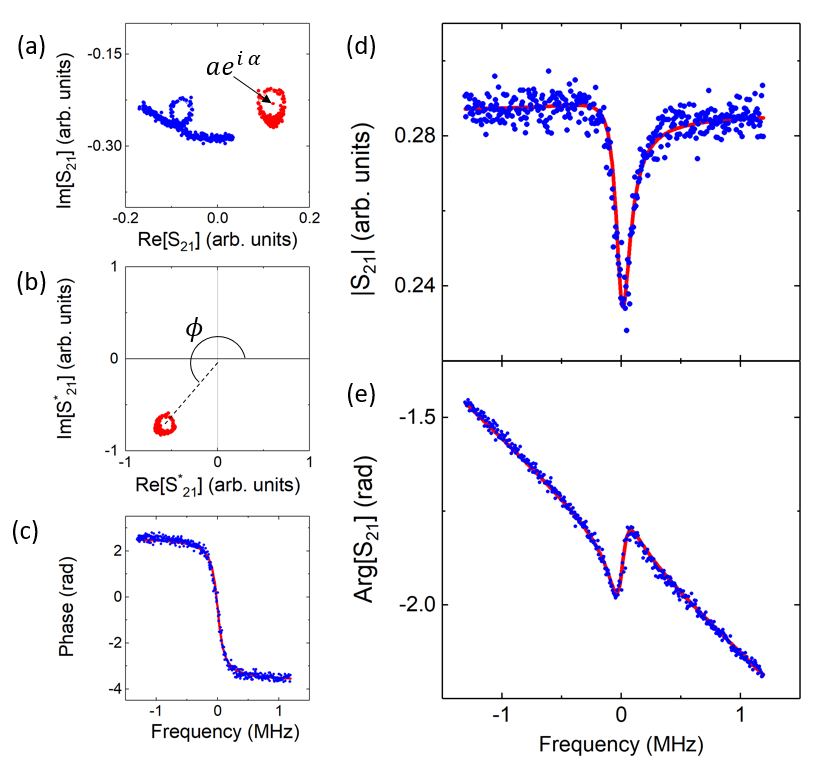}
    \caption{Fitting procedure. (a) Blue and red points correspond to the transmission measured before and after the phase delay is removed, respectively. After removing the phase delay, the data form a circle on the IQ plane with the center at  $ae^{i\alpha}$. (b) Normalized transmission $S_{21}^*$ on the complex plane. The angle between the center of the  $S_{21}^*$ circle and the real axis corresponds to $\phi$. (c) The phase versus frequency (blue points) fitted with $\theta = \theta_0 + \arctan [2Q_l (1-f/f_r)]$ (red curve). (d,e) Measured data (blue points) and the fit with Eq. (\ref{eq:sm1}) (red curve). }
    \label{figS:fitting}
\end{figure}

The magnitude and phase of the transmitted signal $S_{21}$ have been used to extract the quality factors $Q_l$, $Q_C$, and $Q_i$ and the resonance frequency $f_r$. Typically, an asymmetry in the coupling of a resonator to the input and output ports results in deviation of the resonator response from a symmetric Lorentzian function \cite{Khalil2012}. If the coupling between the resonator and the transmission line is weak, the frequency dependence $S_{21}(f)$ near the resonance frequency $f_r$ is described by the following equation \cite{Khalil2012,Probst2015}:
\begin{align}
S_{21}(f)= a \mathrm{e}^{i\alpha}  \mathrm{e}^{-2 i \pi f \tau} \frac{1-(Q_l/|Q_c|\mathrm{e}^{i\phi})}{1+ 2 i Q_l (f/f_r -1)}.   
\label{eq:sm1}
\end{align}
The phase delay $\tau$ can be found from the value of $d[\text{Arg}(S_{21})]/df$ measured over a range of $f$ away from the resonance. All other parameters in Eq. (\ref{eq:sm1}) have been determined similar to the iteration procedure described in Ref. \cite{Probst2015}. We first selected the initial values of unknown parameters  in Eq. (\ref{eq:sm1}), and ran a multi-variable nonlinear fitting procedure for the entire model. The output of the nonlinear fit was used to obtain the final values of unknown parameters and the error bars. The parameter initialization procedure was as follows. After elimination of the phase delay $\mathrm{e}^{-2 i \pi f \tau}$, the data $S_{21} (f)$ formed  a circle on the IQ-plane [Fig. \ref{figS:fitting}(a)]. The prefactor $a\mathrm{e}^{i\alpha}$ corresponds to the center of this circle. For the normalized circle  $S_{21}^* = S_{21} (f)/a\mathrm{e}^{i\alpha -2i \pi f \tau}$, the angle between the off-resonance points and the $I$-axis corresponds to $\phi$, and the circle diameter corresponds to the ratio of $Q_l/|Q_c|$ [Fig. \ref{figS:fitting}(b)]. Next we translated $S_{21}^*$ so that the circle center  coincided with the origin.   $Q_l$ can then be obtained from fitting the phase of the translated $S_{21}^*$, $\theta$, versus frequency with $\theta = \theta_0 + \arctan [2Q_l (1-f/f_r)]$ [see Fig. \ref{figS:fitting}(c)]. Figures \ref{figS:fitting}(d,e) show the experimental data with the result of fitting.

\section{Critical currents of narrow disordered films}
To calculate the Ginzburg-Landau depairing current $I_{dp}(0)$ for strongly disordered Al films at $T \ll T_C$, we used the equation for the critical supercurrent density $j_c=2/(3\sqrt{3})( e \hbar n_s)/(m_e \xi)$ \cite{Tinkham2004}. The concentration of Cooper pairs $n_s$ can be found either from the measured kinetic inductance per square $L_{K\Box}$, or from the result of the Mattis-Bardeen theory $L_{K\Box}=m_e/(2e^2 n_s t)=(\hbar R_{\Box})/\pi \Delta$ where $t$ is the film thickness. The supercurrent density is uniform over the cross section of a superconducting film provided that the film width $W<\lambda^2/t$, where $\lambda$ is the London penetration length. This condition is satisfied for all studied films. Thus, one can estimate $I_c$ as
\begin{equation}
\begin{aligned}
    I_{dp}(0) &=j_c \cdot (Wt) \\
    &=\frac{1}{3\sqrt{3}} \frac{\pi \Delta}{e R_{\Box} \xi(0) } W \\
    &\approx 1.07 \frac{k_B T_c}{e R_\Box \xi(0) } W .
\end{aligned}
\label{eqs:I_dp}
\end{equation}

The coherence length $\xi(0)$ can be found from the data on the upper critical magnetic field for granular Al films, $B_{C2}\approx 4T$ \cite{Cohen1968,Dynes1981}. This yields an estimate $\xi(0)=\sqrt{\Phi_0/ (2\pi B_{C2})} \approx 10$ nm. The data in Table \ref{tabS:table_1} show that the values of the microwave current $I_*=\sqrt{2P_*/Z}$, which corresponds to the onset of strong nonlinearity of the resonator response, are of the same order of magnitude as the current $I_{dp} (0)/\sqrt{Q_l}$ corresponding to the bifurcation threshold.

\begin{table}[htp]
\begin{centering}
\caption{Summary of predicted and measured depairing currents}
\label{tabS:table_1} 
\par\end{centering}
\begin{tabular}{ccccccc}
        \hline
        \hline
         \multirow{2}{*}{\#} & \multirow{2}{*}{$Q_l \cdot 10^4$} & $\rho$ & W  & $I_*$ & $I_{dp}(0)$ & \multirow{2}{*}{$\frac{I_*}{I_{dp}/\sqrt{Q_l}}$} \\
         
         \ & & m$\Omega \cdot$cm & $\mu$m & \text{nA} & $\mu$A 
         \\
        \hline
        1& 3.7& 19.2& 11& 90& 110.2&0.16\\
        \hline
        2& 1.9& 4.2& 7.4& 240& 52.9&0.63\\
        \hline
        3& 3.9& 4.2& 1.3& 30& 12.6&0.47\\
        \hline
        4& 8.4& 9.9& 0.8& 20& 4.6&1.26\\
        \hline
        \hline
    \end{tabular}
\end{table}

\section{Details of $f_r (T)$ and $Q_i (T)$ fitting}
To identify the dominant mechanisms of losses in the studied resonators, we have analyzed the experimental dependences $f_r (T)$ and $Q_i (T)$ on the basis of the theory of two-level systems \cite{Phillips1987} and the Mattis-Bardeen theory of the complex impedance of superconductors \cite{Mattis1958}. 

The losses due to the real part of the complex impedance of superconductors, $\sigma = \sigma_1-i\sigma_2$, can be estimated using the Mattis-Bardeen theory. In the thin film limit \cite{Gao2008}:
\begin{align*}
    \delta_{MB} (T)=\sigma_1 (T)/\sigma_2(T),
\end{align*}
where
\begin{equation}
\begin{aligned}
    \sigma_1( T) = 
    & \frac{\sigma_n}{h \nu} \int \limits_\Delta^\infty [ f(E) - f(E+h \nu) ] \times \\
    & \frac{ E^2 + \Delta^2 + E h \nu }{ \sqrt{E^2 - \Delta^2} \sqrt{(E+h \nu )^2 - \Delta^2} } dE  ,\\
\end{aligned}
\end{equation}
\begin{equation}
\begin{aligned}
 \sigma_2( T) = 
    &\frac{\sigma_n}{h \nu} \int \limits_{\Delta-h\nu}^\Delta [ 1 - 2f(E+h \nu) ] \times \\
    & \frac{E^2 + \Delta^2 + E h \nu }{\sqrt{E^2 - \Delta^2} \sqrt{(E+h \nu )^2 - \Delta^2} } dE  ,\\    
\end{aligned}
\end{equation}
 $f(E)$ is Fermi-Dirac distribution function, $\Delta(0)$ is the energy gap. The temperature-dependent shift of the resonance frequency $f_r$ is given in the main text by Eq. (3). Since the frequency shift $df_r (T)$ does not depend on the MW power, the fitting procedure included the following steps:
\begin{itemize}
    \item fitting  the $df_r (T)$ dependence with only two free parameters: $\Delta(0)$ (which controls the behavior of $\delta f_r^{MB} (T)$ term at $T>300$ mK), and the product of the participation ratio and the material loss tangent, $V_f\delta_0$ (which governs the rising part of $df_r (T)$);
    \item finding the index $\beta$ from the linear part of ${Q_i (\bar{n})}$ at $T = 25$mK plotted on the double-log scale.
    \item knowing $V_f\delta_0, \beta, \Delta(0)$, one can find the low-temperature characteristic power (i.e. the number of photons $n_c{(0)}$) using ${Q_i (\bar{n})}$ measured at the base temperature $T = 25$mK;
	\item finding  $n_c{(T)}$  by fitting $Q_i (T)$ data for both low and high values of the input power.
\end{itemize}

\begin{figure}[h]
    \centering
    \includegraphics[width=\linewidth]{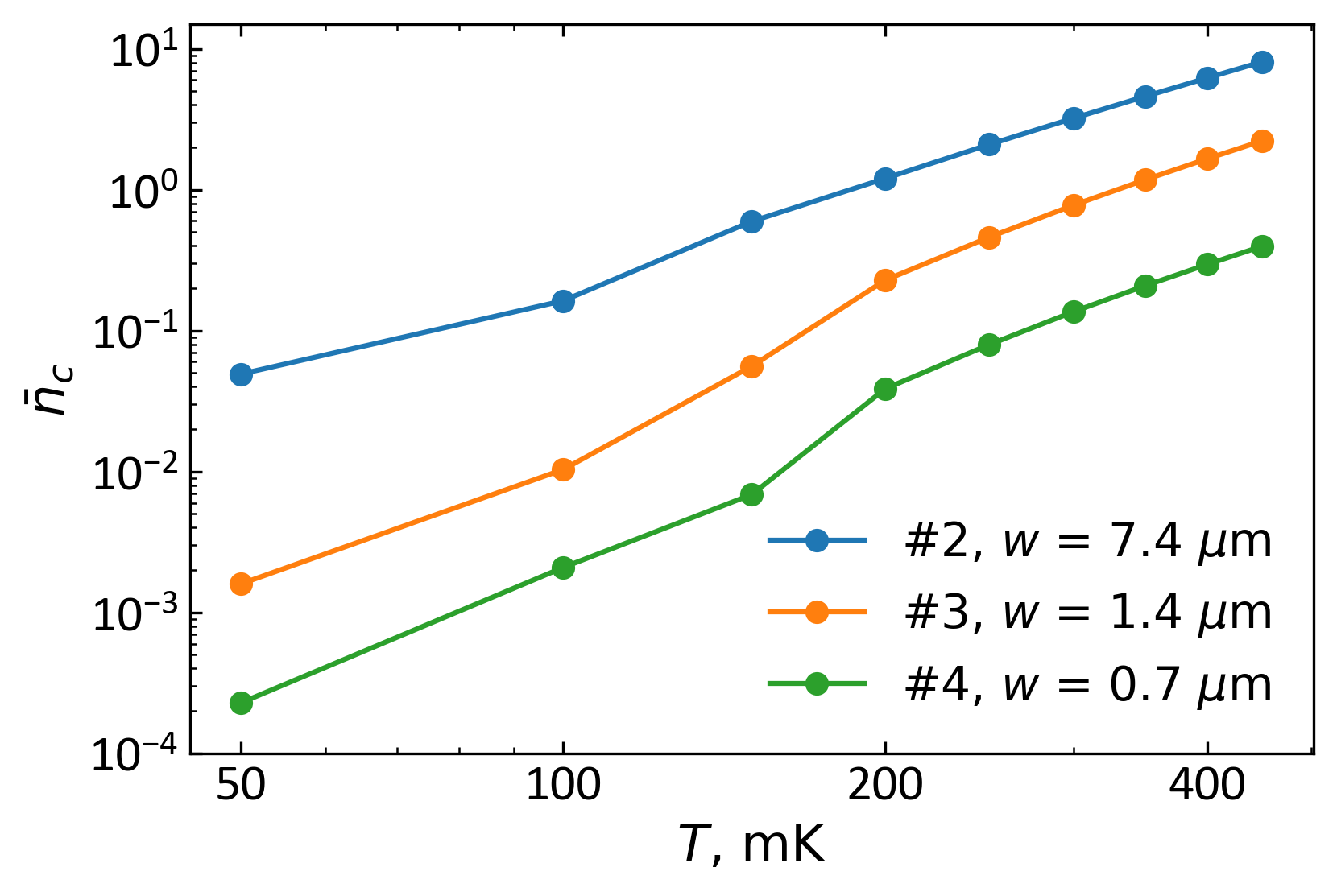}
    \caption{The temperature dependences of $n_c$ for different resonators.}
    \label{figs:nc_T}
\end{figure}

Significant change in the population of the ground and excited TLS states due to Rabi oscillations is expected at the average number of photons in the resonator $n > n_c$. The characteristic value $n_c \sim 1/\sqrt{\tau_1 \tau_\phi }$ depends on the TLS relaxation time $\tau_1(T)$ and the dephasing time $\tau_\varphi (T)$. In agreement with Ref. \cite{Lisenfeld2010}, where the TLS relaxation time was shown to be $\tau_{1,2}\approx 
\tau(1 + \gamma T^\xi)$, we found that the extracted characteristic values of $n_c$ depend on the temperature as $n_c(T)=n_c(0)  +\mu T^\alpha, \alpha \approx 2$ (Fig. \ref{figs:nc_T}).

\section{Scaling of $P_c(0)$}
The ground and excited states of TLS become equally populated  when the Rabi oscillation frequency $\Omega=(d\cdot E)/\hbar$ exceeds the rate  $1/\sqrt{\tau_1 \tau_\phi }$, or, equivalently,  when the electric field in the TLS-occupied volume exceeds the critical value $E_c \approx \hbar /(d\sqrt{\tau_1 \tau_\phi})$. In order to understand the variation of the observed characteristic power for different resonators, we considered the dependence of the maximum electric field near the surface on the resonator parameters. 

The standard way to evaluate the characteristics of CPW resonators is by the Schwarz-Cristoffel (SC) mapping of the coplanar topology to the trivial parallel-plate capacitor geometry. Let us consider a zero-thickness CPW with a central strip width $2a$ and a ground-to-ground distance $2b$. The transfer function for the mapping of the upper half-plane to the rectangle is given by 
\begin{align}
    \xi(w)=A\int \limits_{0}^{w} \frac{dw'}{(w'-a)(w'+a)(w'-b)(w'+b) }.
\end{align}

Here $A$ is an integration constant, chosen to be $A=1$. The half-width of the equivalent capacitor are calculated as
\begin{equation}
\begin{aligned}
    \alpha =\xi(a)=\frac{1}{b} \int \limits_{0}^{1} \frac{dt}{\sqrt{(1-t^2 )(1-t^2 k^2)}} \equiv   \frac{1}{b} K(k)   .
\end{aligned}
\end{equation}
$K(k)$ is also known as the complete elliptic integral of the first kind, $k=a/b$. Similarly, the height of the capacitor is
\begin{equation}
\begin{aligned}
    \beta =\frac{1}{b} \int \limits_{0}^{1/k} \frac{dt}{\sqrt{(1-t^2 )(1-t^2 k^2)}} \equiv   \frac{1}{b} K(\sqrt{1-k^2})   .
\end{aligned}
\end{equation}
The electric field in the $\xi$-plane for the given voltage $V$ across the capacitor is uniform and can be easily obtained as 
\begin{align}
    E_\xi=\frac{V}{\beta}=\frac{V}{K(\sqrt{1-k^2})} b    .
\end{align}
The corresponding field in the $w$-plane scales with the factor $\xi'(w)=d\xi/dw$ which is 
\begin{align}
    \xi'(w)= \frac{1}{\sqrt{(a^2-w^2 )(b^2-a^2)}}   .
\end{align}
Thus, for example, the field strength at the center of microstrip is
\begin{align}
    |E_w(w=0)|= E_\xi \cdot \xi'(w=0)=\frac{V}{K(\sqrt{1-k^2})a} .
\end{align}
Accordingly, the power in the CPW can be written as
\begin{align}
    P \sim \frac{V^2}{Z} \sim \frac{E^2}{Z} a^2 K(\sqrt{1-k^2})^2.    
\end{align}
Therefore, we expect that the characteristic power scales as $P_c \sim (a^2 K(k')^2)/Z$, which is in agreement with the experimental data.

\begin{figure}[hb]
    \centering
    \includegraphics[width=\linewidth]{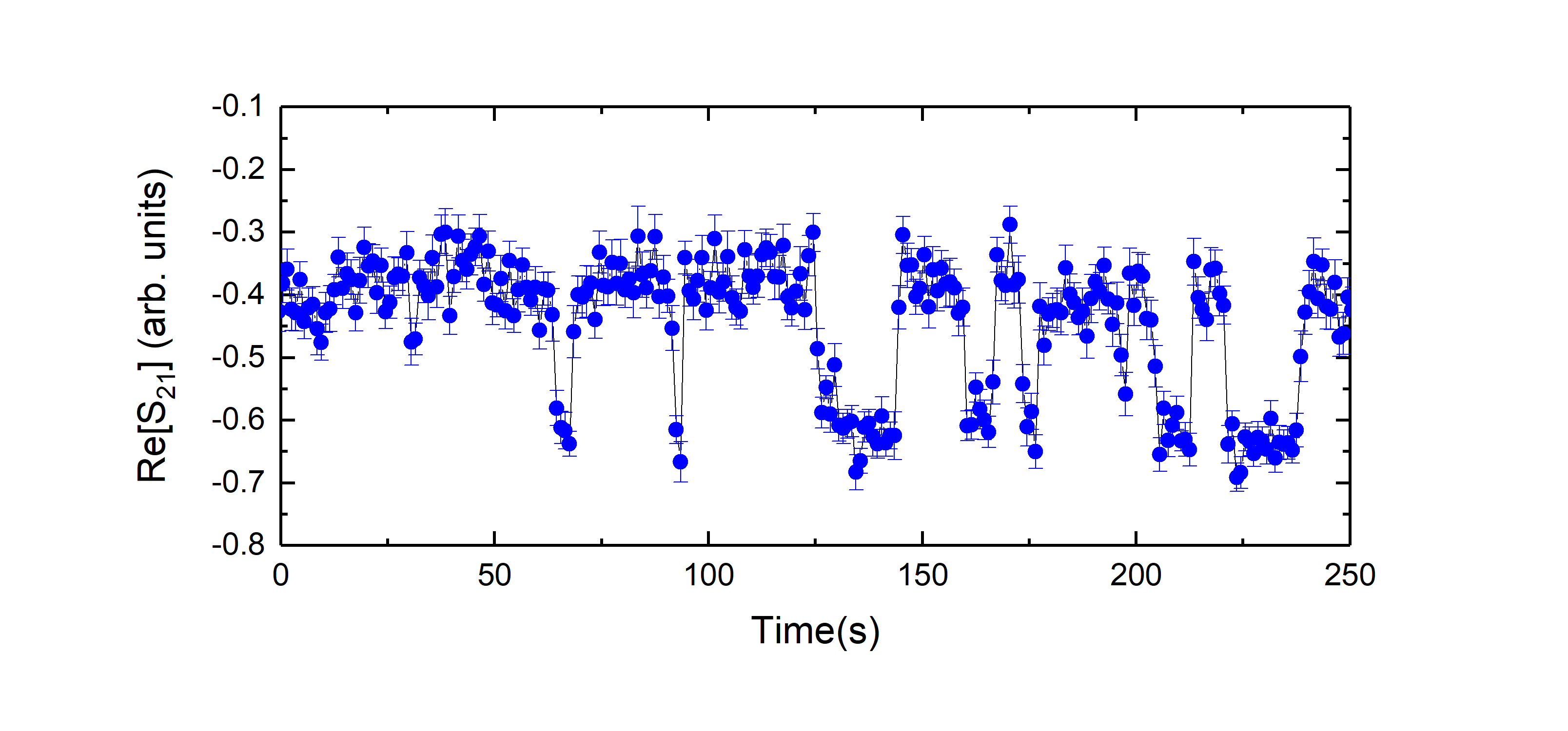}
    \caption{The time dependence of Re[$S_{21}$] measured at $T=25~mK$ at a fixed frequency on the slope of a resonance dip. The microwave power corresponds to $\langle n\rangle\sim 1000$. Each point corresponds to the data averaging over 1 sec.}
    \label{figs:telegraph_noise}
\end{figure}

\section{Telegraph Noise in the resonators }
Interactions between the high-frequency (coherent, $E>k_B T$) TLS with the low-frequency (thermal, $E<k_B T$) fluctuators result in the TLS spectral diffusion as well as the flicker noise.  The telegraph noise in the resonance frequency $f_r$ is expected if some of the TLS with $f\approx f_r$ are strongly coupled to a resonator.
Typical TLS densities for Al/AlOx junctions are $\sim$1~(GHz$\cdot\mu m^2)^{-1}$ \cite{Muller2017}. Interestingly, the number of strongly coupled TLS for our resonators (assuming that the strongly coupled TLS are in the oxide layer of the resonator) is of the order of unity [$1~(GHz\cdot \mu m^2)^{-1}\times 0.1\text{MHz}\times 10^4 \mu m^2$]. To study the telegraphy noise, we repetitively measured $S_{21}$ at a fixed frequency on a slope of the resonance dip for a few minutes. Figure \ref{figs:telegraph_noise} shows an example of the measured telegraph noise in Re[$S_{21}$]. The characteristic time scale of random switching between two Re[$S_{21}$] levels is 1-10 seconds. 


\begin{figure}[t]
    \centering
    \includegraphics[width=0.8\linewidth]{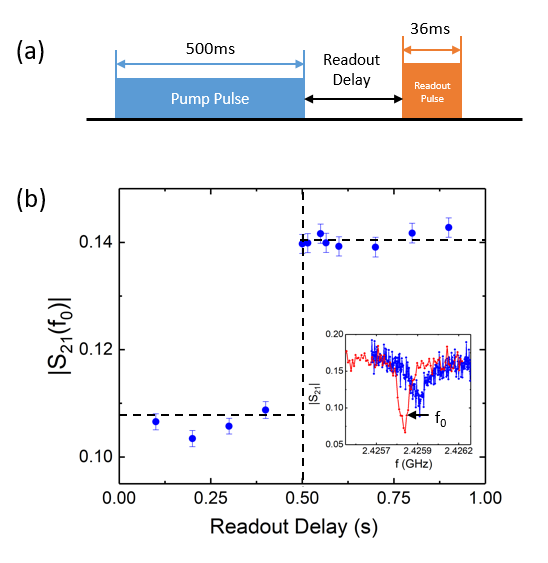}
    \caption{(a) The pulse sequence. (b) The time dependence of $|S_{21}|$ measured at $f_0 = 2.4258$ GHz. The pump pulse at $f_p=f_0+1$ MHz was applied between $t=0~s$ and $t=0.5~s$. The pump tone power corresponds to $\bar{ n}_p \approx 1000$. Each data point was averaged over 4000 cycles with the same readout delay time. The inset shows CW measurement of $S_{21}$ versus $f$ with (red) and without (blue) the pump signal and indicates the position of $f_0$ used in the relaxation time measurement. The readout power was at the single photon level for all measurements on this plot.}
    \label{fig:tls_relaxation}
\end{figure}
\section{Pump-probe measurements of the TLS relaxation time}
We have performed the time domain measurements of the TLS relaxation time for resonator \#1 using the pulse sequence shown in Fig. \ref{fig:tls_relaxation}(a). A 0.5 s-long pump pulse was applied to the resonator at the beginning of each duty cycle. A readout pulse at the single-photon power level lasting for 36 ms followed the pump pulse and was digitized to obtain $S_{21}$. The readout delay time was varied between 0 s and 1 s. Figure \ref{fig:tls_relaxation}(b) shows the result of the experiment at the readout frequency $f_0=2.4258~$GHz and the pump frequency $f_p=f_0+1$ MHz.  The change in $|S_{21}(f_0)|$ at $t=0.5~s$ is consistent with CW measurements at the same readout frequency and power level when a pump tone was turned on and off. This indicates that an upper limit of the TLS relaxation time for our sample is much less than $36$ ms.

\end{document}